\documentclass[aps,endfloat,floatfix,showpacs]{revtex4-1}
\newcommand{\be}{\begin{equation}}
\newcommand{\ee}{\end{equation}}
\newcommand{\bea}{\begin{eqnarray}}
\newcommand{\eea}{\end{eqnarray}}

\newcommand*\Bell{\ensuremath{\boldsymbol\ell}}
\newcommand{\tj}[6]{ \begin{pmatrix}
   #1 & #2 & #3 \\
   #4 & #5 & #6 
\end{pmatrix}}

\usepackage{graphicx}% Include figure files
\usepackage{color}
\usepackage{amsmath}

\begin{document}

\title{Quasi-1D ultracold rigid-rotor collisions : reactive and non-reactive cases}
\author{R. Vexiau}
\affiliation{
Laboratoire Aim{\'e} Cotton, CNRS, Universit{\'e} Paris Sud, ENS Cachan, Universit{\'e} Paris Saclay, 91405 Orsay Cedex, France
}
\author{J.-M. Launay}
\affiliation{
Univ Rennes, CNRS, IPR (Institut de Physique de Rennes)-UMR 6251, F-35000 Rennes, France
}
\author{A. Simoni}
\affiliation{
Univ Rennes, CNRS, IPR (Institut de Physique de Rennes)-UMR 6251, F-35000 Rennes, France
}

\date{\today}

\begin{abstract}

We study polar alkali dimer scattering in a quasi-1D geometry for
both reactive and non-reactive species. Elastic and reactive rates are
computed as a function of the amplitude of a static electric field within
a purely long-range model with suitable boundary conditions at shorter
range. We describe the diatomic molecules as rigid rotors and results are
compared to the fixed-dipole approximation. We show in particular that
for molecules with a sufficiently strong induced dipole moment oriented
perpendicular to the trap axis, the long-range repulsive interaction
leads to the suppression of short-range processes. Such shielding effect
occurs for both reactive and non-reactive molecules, preventing two-body
reactions as well as losses due to complex-mediated processes [Phys. Rev. A
{\bf 85}, 062712 (2012)] from occurring. The present results demonstrate
the possibility to suppress loss rates in current ultracold molecule
experiments using 1D confinement.

\end{abstract}

\maketitle

%\font\smallfont=cmr7

\section{Introduction}

Ultracold molecules have added a new twist to the field of cold
atoms~\cite{LDC-2009-NJP-055409,Quemener2012}. First groundbreaking
experiments were carried out several years ago at JILA, where a
dense gas of ultracold fermionic KRb dimers was produced by two photon
association~\cite{2008-KKN-SCI-322}. This molecular species is reactive, a
feature that has been explored in the context of quantum-state controlled
chemical reactions~\cite{2010-SO-SCI-853}. However, from the standpoint
of many-body studies the reactivity of KRb is in general a drawback
leading to fast particle loss from the trap.

More recently, different groups have
reported the production of {\it non-reactive} ultracold alkali
dimers~\cite{2012-TT-PRL-205301,2014-PKM-PRL-255301,2015-JWP-PRL-205302,2016-MG-PRL-205303,2017-TMR-PRL-143001,2018-Ye-eaaq0083}.
Such molecules, if prepared in the absolute single-particle
energy state, are strictly stable under two-body
collisions. However, there is experimental evidence that even
under these favorable conditions inelastic losses still occur at fast
pace~\cite{2014-PKM-PRL-255301,2015-JWP-PRL-205302,2016-MG-PRL-205303,2018-Ye-eaaq0083}.
This is true in particular for bosonic molecules, whereas fermionic
samples appear to be more stable~\cite{2017-TMR-PRL-143001}. A mechanism
that might explain the observed losses is related to the existence
of Fano-Feshbach resonances available with high density at thermal
energies even in an ultracold gas. Presence of such resonances increases
the collision lifetime of a pair of molecules, creating a tetrameric
complex. Such metastable complex can be lost from the trap since it is not
necessarily trapped in the optical lattice and can undergo recombination
to deeper levels by colliding with a third molecule, a process often
referred to as ``sticky collision''~\cite{2012-MM-PRA-062712,2013-MM-PRA-012709}.
Excitation of the complex by the trapping laser has also been recently
proposed as an alternative mechanism that might explain the observed
loss rates~\cite{2019arXiv190506846C}.

In order to increase the sample lifetime, it is therefore important
to devise strategies to prevent molecules from approaching at
short distances, where detrimental inelastic processes take
place. Unfortunately, in three dimensions there always exists
an attractive head-to-tail reaction path leading to the short
range. This circumstance has been shown to result in a strong
dependence $\sim d^6$ of the reactive rate on the induced dipole
moment $d$~\cite{GQ-2010-PRA-022702}. More subtle quantum effects
have been proposed to control the reaction dynamics of alkali dimers
prepared in rotationally excited states~\cite{2015-GW-NJP-035015}.
Microwave shielding leading to a dramatic lifetime increase has also been
recently theoretically studied~\cite{2018-LL-PRL-163402,2018-TR-PRL-163401}.

The situation is different if molecules are confined in tight
traps, for instance by optical means. In fact, confinement can then
be easily designed in such a way that molecule pairs will tend
to collide in a {\it repulsive} side-to-side configuration, the
head-to-tail pathway being energetically unfavored by the confining
potential. This approach has been demonstrated both experimentally
and theoretically to lead to a drastic increase of the lifetime of
reactive polar molecules trapped in one and two spatial dimensions
~\cite{AM-2010-PRL-073202,2011-MHGDM-NP-502,2012-AC-PRL-080405,2015-AS-NJP-013020}.

Our previous study of reactive collisions in one-dimensional optical tubes
has addressed highly reactive species~\cite{2015-AS-NJP-013020}. Therein,
polar molecules have been simply described as fixed dipoles of magnitude
equal to the induced dipole moment. Resonances have been found to be
strongly quenched by inelastic processes and have no influence on the
scattering cross sections. Aim of the present work is twofold. First
we relax the fixed-dipole approximation by introducing explicitly
the rotational degrees of freedom of the colliding diatoms. Secondly,
in addition to studying reactive collisions we introduce a model for
non-reactive molecules, in which scattering cross sections can indeed
be dominated by dense resonance spectra. We compare both reactive and
non-reactive rigid-rotor models with the fixed-dipole approximation. As
a main result, we show that resonances in elastic cross sections and thus
possibly complex-mediated collisions can be suppressed using the dipole-dipole
repulsion induced by an applied electric field.

The paper is organized as follows. Section~\ref{formalism} introduces
the formalism and our numerical approach. Section~\ref{c6} presents
the long-range multipolar expansion of the molecule-molecule
interaction. Results for different dialkali species and collision
energies are presented as a function of an applied electric field in
Sec.~\ref{results}. A short conclusion summarizes this work.

\section{Formalism}
\label{formalism}

We consider two molecules A and B confined in a quasi one-dimensional
geometry by a potential approximated as an harmonic trap. The quadratic
nature of the confining potential allows one to separate center-of-mass
and relative motion. The rigid-rotor Hamiltonian describing the
collision in relative coordinates includes the kinetic energy, the
confinement potential, and an interaction term $V_{\rm int}$ defined
below in Eq.~\eqref{vint} which includes inter-molecular forces and the
Stark interaction energy with the external electric field. In this work
we will describe the dimer molecules as rigid rotors, an approximation
valid as long as the intermolecular distance remains large compared to
the extent of the internal coordinates.

The kinetic energy of the dimer+dimer complex will be decomposed
into a sum of three independent terms depending respectively on
the relative distance $\mathbf R$ between the two centers of mass,
the internal coordinate of the dimer A and the internal coordinate
of dimer B~\cite{Avoird1994}. Furthermore, we will assume that the
dimers remain in their vibrational ground level reducing the dimer
contribution to the kinetic energy to a purely rotational term. 

Putting all terms together, the total Hamiltonian reads :
\be
\label{hamiltonian}
H = T_R + T_{\rm rot} + V_{\rm trap} + V_{\rm int} = -\dfrac{\hbar^2}{2
\mu} \dfrac{d^2}{dR^2} + \dfrac{{\Bell}^2}{2\mu R^2} +
\dfrac{1}{2}\mu\omega_{\perp}^2\rho^2 + B_v\mathbf{J}^2_A +
B_v\mathbf{J}^2_B + V_{\rm int}(\mathbf{R}) .
\ee
Here $\Bell$ is the orbital angular momentum of the relative motion,
$\mathbf{J}_A$ and $\mathbf{J}_B$ are the angular momenta of A and B with
rotational constant $B_v$, $\omega_{\perp}=2\pi\nu_{\perp}$ is the trap
frequency and $\rho$ the distance to the trap axis $z$.  We use for each
molecule the spectroscopically determined rotational constant $B_v$ of the
ground level $X^1\Sigma^+;v=0$. Values and reference of the spectroscopic
study can be found in~\cite{Lepers2013}.  The transverse oscillator length
characterizing the trap size will be defined as $a_{\rm ho}=\sqrt{\hbar /
(\mu \omega_{\perp})}$. As demonstrated in three spatial dimensions,
hyperfine interactions would greatly increase the complexity of the
problem~\cite{2012-MM-PRA-062712,2013-MM-PRA-012709}. However, we do not
expect such added complexity to bring novel qualitative features to the
main effects we wish to demonstrate and hyperfine interactions will be
neglected in this work.

We will now extend to rigid rotors the computational approach developed
for fixed dipoles in our previous work~\cite{2015-AS-NJP-013020}. Our
method to construct the scattering wavefunction consists in a simultaneous
expansion of the angular part of the solution in a suitable internal basis
whereas the radial coordinate is discretized on a grid of points. The
radial discretization is detailed in Sec.~\ref{radial}. The angular basis
comprises the orbital angular coordinate $\hat{R}$ as well as any internal
degree of freedom of the diatomic, as described in Sec.~\ref{basis}.

\subsection{Radial discretization}
\label{radial}

The radial discretization is performed according to the spectral element
approach~\cite{Kerniadakis,2017-AS-JCP-244106}. Briefly, we choose a
minimum $r_c$ and a maximum $r_{\rm max}$ radial distance and define
a solution interval $I = \left[r_{c}, r_{\rm max} \right]$ which is in
turn partitioned into a set of N non-overlapping sectors. A number of grid
points and a basis set of Gauss-Lobatto cardinal functions associated with
those points are assigned to each sector. The wavefunction is represented
on the discrete basis and continuity of the wavefunction and of its radial
derivative is enforced at the connection points between sectors. This
strategy results in a highly sparse matrix in the grid indices. It should
be remarked that our method allows different angular bases to be used in
each sector. With the basis over all coordinates defined, we can rewrite
the Schr\"odinger equation as a linear system that can be solved for a
matrix solution $\Psi$ in the interval $I$ provided boundary conditions
are assigned at the endpoints $r_c$ and $r_{\rm max}$.

At the right endpoint we impose $\Psi^\prime(r_{\rm max}) = \mathbf{ I}$,
the matrix solution at $r_{\rm max}$ becomes then equal to the $R$-matrix
defined as $\mathbf{R} \equiv \Psi \left( \Psi^\prime \right)^{-1}$. From
$\mathbf{R}$ one can then extract the scattering matrix and hence
all physical observables such as the elastic and reactive collision
rates~\cite{AM-2010-PRL-073202,2015-AS-NJP-013020}.

We will impose two different boundary conditions at $r_c$. The
first assumes that at short distance the molecules react with unit
probability. We thus require that the spherical surface $R=r_c$
is totally absorbing , {\it i.e.} across the surface we only have
incoming flux and no reflected outgoing flux. In practice we first define
local adiabatic channels $| \alpha \rangle$ and corresponding energies
$E_\alpha$ by diagonalizing the angular Hamiltonian $ T_{\rm rot}+ V_{\rm
int} $ at distance $R=r_c$. Next, we assume that the wavefunction
can be described by a pure incoming spherical wave in each channel
$\alpha$ for $R \simeq r_c$. The logarithmic derivative $Z(r_c) \equiv
\Psi^\prime(r_c) \Psi^{-1}(r_c)$ is therefore diagonal with elements
$( -i k_{\alpha} - \dfrac{1}{2}k_{\alpha}'k_{\alpha}^{-1/2} )$,
where $k_{\alpha} = \sqrt{2 \mu (E_{\rm coll} -E_\alpha) / \hbar^2}$
is the channel wave vector and the derivative is taken with respect to
$R$. This method has been shown to give accurate prediction for dialkali
reactive species~\cite{ZI-2010-PRL-113202}.

Our second approach amounts to the Dirichlet boundary condition
$\Psi={\bf 0}$, {\it i.e.} the wavefunction is required to have
a nodal surface for $R=r_c$. In this description we can observe
resonances between the incoming open channel with collision energy
$E_{\rm coll}$ and bound levels of closed channels with energy close
to $E_{\rm coll}$. Since the radius $r_c$ is chosen arbitrarily and
there is {\it a-priori} no reason for the wavefunction to have a nodal
surface for $R=r_c$, we cannot predict the location of the resonances in
terms of the amplitude of the electric field. As a matter of fact, the
short-range collision parameters (more precisely, the quantum defects,
see~\cite{2012-MM-PRA-062712,2013-MM-PRA-012709}) can be considered as
stochastic variables arising from the extremely complex four-body dynamics
taking place inside $r_c$. In this view, our specific Dirichlet condition
can be considered as one possible realization of such complex process.

The specific value $r_c=40~a_0$ has been chosen so to satisfy the
following criteria. Firstly, the rigid rotor model is expected to be
accurate in the external region $R>r_c$. Second, motion is semiclassical
near $r_c$ for the adsorbing model of Ref.~\cite{ZI-2010-PRL-113202}
to apply. Finally, since we are interested in resonance spectra,
the density of states near threshold must be the same for the Dirichlet
``truncated potential'' as for the full one. We have checked that for
our choice of $r_c$ this holds true up to energies on the order of 1~K
in the case of a single deep potential well with van der Waals tail.

The present approach can therefore expected to give useful insight into
the collisions dynamics and it allow us to compare on an equal footing
different approaches, namely the fixed-dipole and the rigid-rotor
descriptions. Finally, note that for the sake of comparison we will use the
non-reactive Dirichlet condition even for intrinsically reactive species
such as LiCs or LiRb.

\subsection{Angular basis}
\label{basis}

The angular part of the wavefunction in the rigid-rotor model
is expanded in the so-called decoupled basis $|J_A M_A,J_B M_B,l
M_l\rangle$, representation defined by the operators $\mathbf{J}^2_A$,
$\mathbf{J}^2_B$, ${\Bell}^2$ with eigenvalues $\hbar^2J_A(J_A+1)$,
$\hbar^2 J_B(J_B+1)$, and $\hbar^2 l(l+1)$ as well as by their projections
on the laboratory axis with eigenvalues $M_A$, $M_B$ and $M_l$.
The laboratory axis is taken as the axis of the confining optical tube. In
parallel configuration where the electric field axis is aligned along the
laboratory axis, the projection of the total angular momentum $M \equiv
M_A + M_B + M_l$ is conserved, allowing us to work with a smaller basis
set. For non-parallel configurations one has to take into account couplings
between different $M$ due to the electric field.

The interaction potential matrix included in Eq.~\eqref{hamiltonian}
\be
\label{vint}
{\mathbf V}_{\rm int}(R) = \dfrac{\mathbf C_3}{R^3} + \dfrac{\mathbf C_6}{R^6} + {\mathbf V}_{{\rm Stark},A} + {\mathbf V}_{{\rm Stark},B}
\ee
comprises the dipole-dipole interaction ${\mathbf C}_3/R^3$, the van der
Walls interaction ${\mathbf C}_6/R^6$, and the Stark term arising from the
interaction between the molecular dipoles ${\mathbf d}_{A,B}$ and the
electric field. The matrix elements of ${\mathbf C}_3$ and ${\mathbf C}_6$ depend on the
angular momenta $J_A$, $J_B$ and $l$ as well as their projection
on the laboratory axis. They are obtained using Wigner-Eckart
theorem, their expression in the space fixed frame can be found
in~\cite{Avoird1980,Krems2004,Micheli2007,Bohn2009}.
For completeness, we report here the expression of the dipole-dipole interaction
\bea
&&   \langle {J}_A' M_A {J}'_B M_B' l' M_l' | \dfrac{\mathbf C_3}{R^3} | {J}_A M_A {J}_B M_B l M_l \rangle = \nonumber \\
    &-&\sqrt{30} \frac{d^2}{R^3} \sum_{m_1 m_2} (-1)^{M_A'+M_B'+M_l'} 
\sqrt{[{J}_A][{J}_A'][{J}_B][{J}_B'][l][l']}
\tj{1}{1}{2}{m_1}{m_2}{-m_1-m_2} 
\tj{{J}_A'}{1}{{J}_A}{0}{0}{0}
\tj{{J}_A'}{1}{{J}_A}{-M_A'}{m_1}{M_A}   \nonumber \\
&\times& \tj{{J}_B'}{1}{{J}_B}{0}{0}{0}
\tj{{J}_B'}{1}{{J}_B}{-M_B'}{m_2}{M_B}  
\tj{l'}{2}{l}{0}{0}{0}
\tj{l'}{2}{l}{-M_l'}{-m_1-m_2}{M_l},
\eea
where the common abbreviation $[X]=2X+1$ has been used.
The Stark term represents an internal interaction that acts on the degrees
of freedom of the individual molecule. The term for molecule A (resp. B) is therefore diagonal
in the quantum numbers of molecule B (resp. A) as well as in the orbital quantum numbers $l M_l$.
The matrix elements are given explicitly for molecule of label A by
\be
   \langle {J}_A' M_A' | {\mathbf V}_{{\rm Stark},A}  | {J}_A M_A  \rangle
=  - {\cal E} \langle {J}_A' M_A' | d_{z,A}  | {J}_A M_A  \rangle
= {\cal E} d \sqrt{[l][l']} \tj{{J}_A'}{1}{{J}_A}{0}{0}{0} \tj{{J}_A'}{1}{{J}_A}{-M_A'}{0}{M_A}
\ee
in terms of the electric field intensity $\cal E$, a formally identical
equation holding for B. 
The electronic van der Waals interaction is in general
anisotropic~\cite{SK-2010-NJP-073041}. However, we have taken explicitly
into account the anisotropic contribution for the study of KRb+KRb
and found it to give negligible corrections to the computed scattering
observables. Therefore, only the isotropic contribution will be included
in our model.

Few additional points must be taken into consideration. First, since we focus
on collisions of identical diatomics we need to use
symmetrized wavefunctions
\be
|\Psi_{J_A M_A J_B M_B  l M_l}\rangle = \dfrac{|J_A M_A J_B M_B  l M_l \rangle +
(-1)^l| J_B M_B J_A M_A l M_l \rangle}{\sqrt{2(1+\delta_{J_A J_B} \delta_{M_A M_B}  )}} .
\ee
For the perpendicular electric field configuration, the Hamiltonian is
also symmetric with respect to reflection across the plane orthogonal to
the trap axis and containing the origin. We can construct a symmetrized
wavefunction
\be
|\Psi_{J_A M_A J_B M_B  l M_l}^\epsilon  \rangle =
\dfrac{| \Psi_{J_A M_A J_B M_B  l M_l}   \rangle + (-1)^{\epsilon + M_A +M_B +M_l}| \Psi_{J_A -M_A J_B -M_B  l -M_l}
   \rangle}{\sqrt{2(1+  \delta_{M_A 0} \delta_{M_B 0} \delta_{M_l 0} )}} .
\ee 

Lastly, while the decoupled basis is useful in the weak electric
field regime, for higher amplitudes, different J levels are
heavily mixed by the field, the system is better described in
a dressed basis set \cite{Friedrich1999,Avdeenkov2006}. To
obtain this basis we numerically diagonalize for each field
amplitude the diatomic potential $B_v\mathbf{J}^2_{X} +
V_{{\rm Stark},X}$ ($X=A,B$). We thus get the eigenfunctions
$|\widetilde{J}{\widetilde{M}_J}\rangle = \sum\limits_J \alpha_J
|JM_J\rangle$ which are in turn combined into the tetrameric wavefunction
$|\widetilde{J}_A\widetilde{M}_A\widetilde{J}_B\widetilde{M}_BlM_l\rangle$.
In order to compute the total potential matrix we first evaluate the
matrix elements in the decoupled basis and then we numerically evaluate
\be
\langle\widetilde{J}_A'\widetilde{J}_B'l' | H |
\widetilde{J}_A\widetilde{J}_Bl\rangle = \sum\limits_{J_A J_B J_A' J_B'}
\alpha_{J_A}\alpha_{J_B}\alpha_{J_A'}\alpha_{J_B'} \langle J_A'J_B'l' | H | J_AJ_Bl\rangle ,
\ee
where the magnetic quantum numbers have been suppressed for notational convenience.

The rigid-rotor description is only used between $R=40~a_0$, chosen as
the $r_c$ boundary value in this work, and $R=200~a_0$. For intermediate
distances above $R=200~a_0$, dipole-dipole interaction between different
rotational level is weak compared to the energy gap of those level. One
can thus focus on the rotational ground level with $\widetilde{J}_A=0$
and $\widetilde{J}_B=0$, using perturbation theory to take into account
excited levels; see Sec.~\ref{c6}. Spherical harmonics $|l M_l\rangle
$are still used to represent orbital motion.

Finally, in the long-range domain, typically $R > 10^4~a_0$,
the cylindrical confining potential of the 2D-trap overcomes the
molecule-molecule interaction, limiting $\hat{R}$ to a small angular
region around the pole. A development in spherical harmonic becomes
increasingly inefficient and we use cylindrical grid basis functions
projected on the spherical surface that stay localized near the pole as
$R$ varies. As an order of magnitude, few hundreds spherical harmonics
are needed to enforce continuity at the connection point between the
intermediate and the long-range regions.

We will compare the rates obtained with the current approach to results
from our previous model where fixed dipoles were used instead of
rigid-rotors~\cite{2015-AS-NJP-013020}. In the latter, the interaction
between the molecule in its rotational ground level and the Stark field
is described as a point particle with a given induced permanent dipole
moment. All rotational effects factor into a unique $C_6$ long-range
coefficient. The total angular wavefunction is expanded on a basis
of orbital spherical harmonics and the Hamiltonian only contains the
orbital kinetic energy, the confinement potential and the interaction
potential $V_{\rm int}$. Such point-particle model is equivalent to our
intermediate-range rigid-rotor model where we include only the rotational
ground state, though the correction to the $C_6$ coefficient needs to
be taken the same in both models; See Sec.~\ref{c6}.

\section{Induced dipole moment interaction}
\label{c6}

Some care must be taken in the definition of the $C_6$ matrix coefficient in
order to have a correct description of the Van der Waals interaction. Two
different approaches are used in this work.

In the first one, used for the intermediate range domain or for the
point particle model, excited rotational states are not explicitly
included in the basis. Resulting from a second order perturbation
treatment, the $C_6$ coefficient is thus written as a sum over all excited
rotational, vibrational and electronic levels of both molecules. Following
Refs.~\cite{Quemener2011, Lepers2013} the $C_6$ can be decomposed into
the sum of two contributions, a term $C_6^e$ including the electronic
transitions contribution and one $C_6^r$ including the rotational
transition contribution. Pure vibrational transitions as well as cross
terms have been shown to be negligible~\cite{Lepers2013}. Values of the
$C_6^e$ are taken from~\cite{Lepers2013}. The $C_6^r$ term is described
in more detail in Sec.~\ref{sec-c6g}.

In the second approach, used for the short range domain, the rotational
states are explicitly included in the basis. The second order term is
thus coming only from the electronic transitions $C_6^e$. An additional
corrective term is however needed to prevent the appearance of unphysical
levels; See Sec.~\ref{unphysical}. 

As mentioned above, in both cases the $C_6^e$ is taken as a purely
isotropic term.

\subsection{Rotational transition contribution}
\label{sec-c6g}

Due to the competition between the electric field and the rotational
Hamiltonian, the rotational spectrum is heavily perturbed. A field free
$C_6^r$ value would thus not be appropriate~\cite{Julienne2011}.

In more detail the expression of the rotational contribution is 
\be
\label{c6g}
C_6^r = \sum\limits_{(\widetilde{J}_A,\widetilde{J}_B)
\ne (\widetilde{0},\widetilde{0})} \dfrac{|\langle
\widetilde{J}_A,\widetilde{J}_B |\mathbf{d}_A.\mathbf{d}_B|
\widetilde{0},\widetilde{0}\rangle |^2}{\Delta E_A + \Delta E_B}
\ee
$\Delta E_A$ (resp. $\Delta E_B$) being the energy difference between
the ground and the excited $J_A$ (resp. $J_B$) level. Dependence on the
space-fixed projections $M_A$ and $M_B$ of the excited rotational levels
are omitted in the formula for ease of reading. For the field free case
$\widetilde{J}\equiv J$ and the only non-zero transition is from $J=0$
to $J=1$. We then obtain the expression for two identical molecules
\be
C_6^r ({\cal E}=0) = \dfrac{d_{\rm perm}^4}{4B_v}
\ee
with $d_{\rm perm}$ the permanent dipole moment of the diatomic, expressed
in its own molecular frame.

To obtain more accurate results for the collision rate we compute an
improved $C_6^r$ value using Eq.~\eqref{c6g} at each electric field
amplitude. We note that for a non-zero field the heteronuclear
molecules present an induced permanent dipole moment in the
laboratory frame. In this case the contribution of matrix elements
like $\langle \widetilde{0},\widetilde{1} |\mathbf{d}_A.\mathbf{d}_B|
\widetilde{0},\widetilde{0}\rangle$ is non negligible. Those transition
terms with one molecule in the ground level and the other in an excited
one are related to the induction interaction $C_6^{\rm ind}$. While the
remaining transitions involving two rotationally excited molecules added to
the electronic $C_6^e$ form the dispersion coefficient $C_6^{\rm disp}$.

\begin{figure}[p] \begin{center}
\includegraphics[width=0.9\columnwidth]{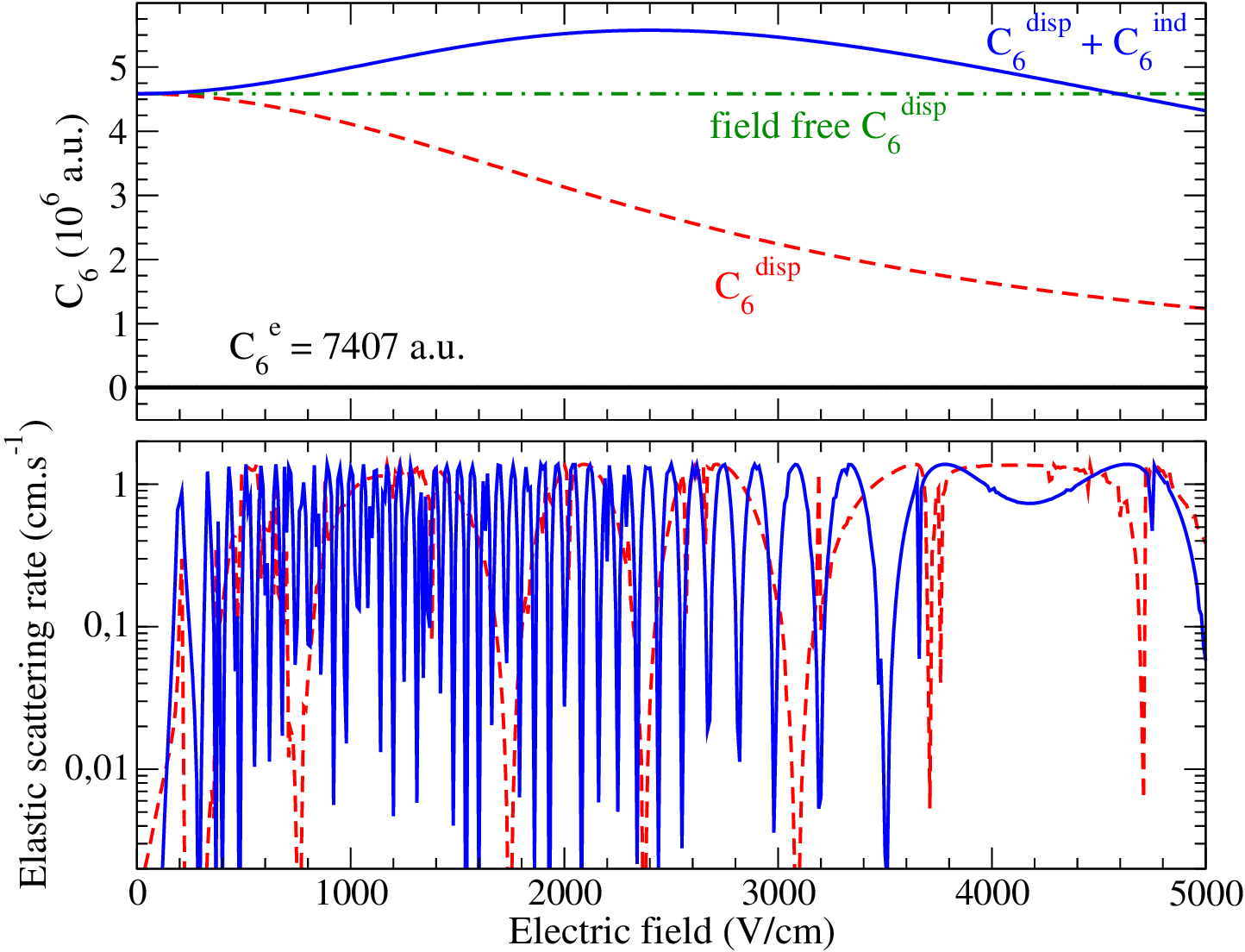} \caption{\label{fig:c6g}
\small (color online) Upper panel : van der Waals $C_6$ coefficient
for LiCs-LiCs interaction. The electric field is parallel to the
trap axis. Four different models were used : an electronic $C_6^e$
(full line at $7407~a.u.$) added to a rotational field-free $C_6^r$
term (dashed line) or a field dependent $C_6^r$ (dot-dashed line)
added to the induction coefficient $C_6^{\rm ind}$ (full line).
Lower panel : elastic collision rate at $E_{\rm coll}=50$~nK imposing the
Dirichlet boundary condition at short range computed for van der Waals
coefficient $C_6^r$ (full line) or $C_6^r+C_6^{\rm ind}$ (dashed line).
} \end{center} \end{figure}

Rotational contributions have been shown to be small for
heteronuclear molecules LiNa and KRb due to their weak permanent dipole
moment~\cite{Lepers2013}.  
However for a molecule like LiCs they are by far the dominant term. In
Fig.~\ref{fig:c6g} we show the difference in $C_6$ coefficients for
LiCs-LiCs interactions depending on the model used. Calculations were
performed with a single $C_6^e$ term, a fixed field-free $C_6$ term,
a field-dependent $C_6^{\rm disp}$ term and finally a field-dependent
$C_6^{\rm disp}+C_6^{\rm ind}$ term. At weak fields the difference
between the values obtained at the different levels of approximation is
small and the collision rate is essentially the same for each model. At
higher fields the dominant term is the $C_3$ dipole-dipole interaction
and modifications of the $C_6$ coefficient have no major impact on
our results. However the $C_6$ interaction is still relevant at short
intermolecular distances. 

Bottom panel in Fig.~\ref{fig:c6g} shows the elastic collision rate
computed using model potentials built with and without the induction
interaction. Calculations are performed at low collision energy with
the Dirichlet boundary condition at short range, such that scattering is
purely elastic. The calculated collision rate presents several resonant
features associated to the presence of quasi-bound states. The origin of
these features will be discussed in more detail in Sec.~\ref{results}.
It is however important to remark here, as apparent from the figure,
that their density at large electric fields is severely underestimated
if one omits the induction interaction.

\subsection{Unphysical states}
\label{unphysical}

When neglecting the rotation in the point-particle model we
still need to compute a corrective $C_6$ term, which means that
the modification of the rotational structure must be taken into
account in the model in second order perturbation theory.  In a
similar fashion, the use of the rigid-rotor basis mentioned in
Sec.~\ref{basis} leads to unphysical states if no correction is
made~\cite{Tscherbul2010,Suleimanov2012}. Indeed due to computational
limitations one has to truncate the basis, neglecting all the levels with
$\widetilde{J} > \widetilde{J}_{max}$. In particular all the coupling
matrix element of the type $\langle \widetilde{J}_A'\widetilde{J}_B'l' |
V | \widetilde{J}_A\widetilde{J}_Bl \rangle $ with either $J_A'$ or $J_B'$
above the angular momenta $\widetilde{J}_{max}$ are cut off. Rotational
levels with $J=\widetilde{J}_{max}$ are the most affected by this
truncation~\cite{Tscherbul2010,Suleimanov2012}. To reduce the effect of
this approximation on the ground state, and thus on the collision rate,
one can simply increase the size of the basis.

Another approach is to use second order perturbation theory. In this
approximation, the omitted couplings are assumed to be a perturbation
to the levels included in our basis, giving rise to a corrective term
$V_{\rm corr}$ to be added to the Hamiltonian. This term is diagonal
with matrix elements
\be
\langle \widetilde{J}_A\widetilde{J}_Bl |
V_{\rm corr} |\widetilde{J}_A\widetilde{J}_Bl\rangle
= \sum\limits_{\widetilde{J}_A'\widetilde{J}_B' > \widetilde{J}}
\dfrac{|\langle \widetilde{J}_A'\widetilde{J}_B'l | V |
\widetilde{J}_A\widetilde{J}_Bl\rangle|^2}{\Delta E_A + \Delta E_B}
\ee
where the sum is over every level above the cutoff angular momentum
and $V$ is the dominant coupling term, the dipole-dipole interaction
$C_3/R^3$ in our case. The square of the matrix element on the rhs leads
to a $R^{-6}$ correction to the $C_6$ coefficient. In this work we take
$\widetilde{J}_{max}$ equal to either 1 or 2, and the sum is carried
out up to $\widetilde{J}=7$. The energy gap $\Delta E$ is taken as the
diatomic energy gap, thus neglecting the molecule-molecule interaction.

\section{Results}
\label{results}

We perform calculations for different bosonic species under various trapping
conditions, collision energy and electric field. Table~\ref{tab:data}
resumes the relevant characteristic physical parameters relevant for
our calculations. The van der Waals length ${\bar a} = (2 \mu C_6 /
\hbar^2)^{1/4}/2$ given in the table represents the average scattering
length for collisions in a pure $C_6/R^6$ potential and can be interpreted
as the range of such potential~\cite{GFG-1993-PRA-546}. We focus
here on an intermediate confinement regime, which we define following
Ref.~\cite{2011-AS-JPB-235201} as $a_{\rm ho} \approx 10~\bar{a}$. The
collision energy of the identical heteronuclear molecules will be fixed
in most calculations to $50$~nK. With reference to the table one can
remark that for a heavy molecule such as LiCs such collision energy can
be considered as ``hot'' in terms of the trap level spacing. In fact,
the gap with the first excited energy level of the transverse harmonic oscillator is
only slightly larger than twice the collision energy.

\begin{table*}[htb]
\begin{tabular}{|c|c|c|c|c|c|c|}
\hline
 & $\bar{a}(a_0)$ & $a_{\rm ho}(a_0)$ & $\nu_{\perp}$(kHz) & $E_{\rm coll} /\hbar \omega_{\rm perp}$ & $d_{\rm perm}$(a.u.)& $B_v$(GHz)\cite{Lepers2013} \\
\hline
$^{39}$K$^{87}$Rb & 117 & 955 & 10 & 1.6[-2] & 0.242 & 1.13\\
\hline
% $^{7}$Li$^{23}$Na & 57 & 619 & 100 & 1.6[-3] & 0.223 & 11.21 \\
% \hline
$^{7}$Li$^{39}$K & 224 & 2236 & 5 & 3.3[-2] & 1.410 & 7.69\\
\hline
$^{7}$Li$^{87}$Rb & 325 & 2473 & 2 & 8.3[-2] & 1.645 & 6.46\\
strong & - & 341 & 100 & 1.6[-3]  & - & \\
\hline
$^{7}$Li$^{133}$Cs & 497 & 6408 & 0.2 & 8.3[-1] & 2.201 & 5.62\\
\hline
$^{23}$Na$^{87}$Rb & 355 & 2286 & 2 & 8.3[-2] & 1.304 & 2.09\\
\hline
\end{tabular}
\caption{\label{tab:data} 
Relevant numerical parameters used in our calculation, as defined in
the text. The reference collision energy $E_{\rm coll}$ is 50~nK. Two
confinement strengths, intermediate and strong, are indicated for the
LiRb dimer.  
}
\end{table*}

We first consider the configuration where the electric field is
parallel to the confinement axis. In this case, molecules tend to
be in attractive head-to-tail configuration and to react at the
cutoff radius $r_c$. 

\begin{figure}[p]
\begin{center}
\resizebox{1.0\columnwidth}{!}{\includegraphics{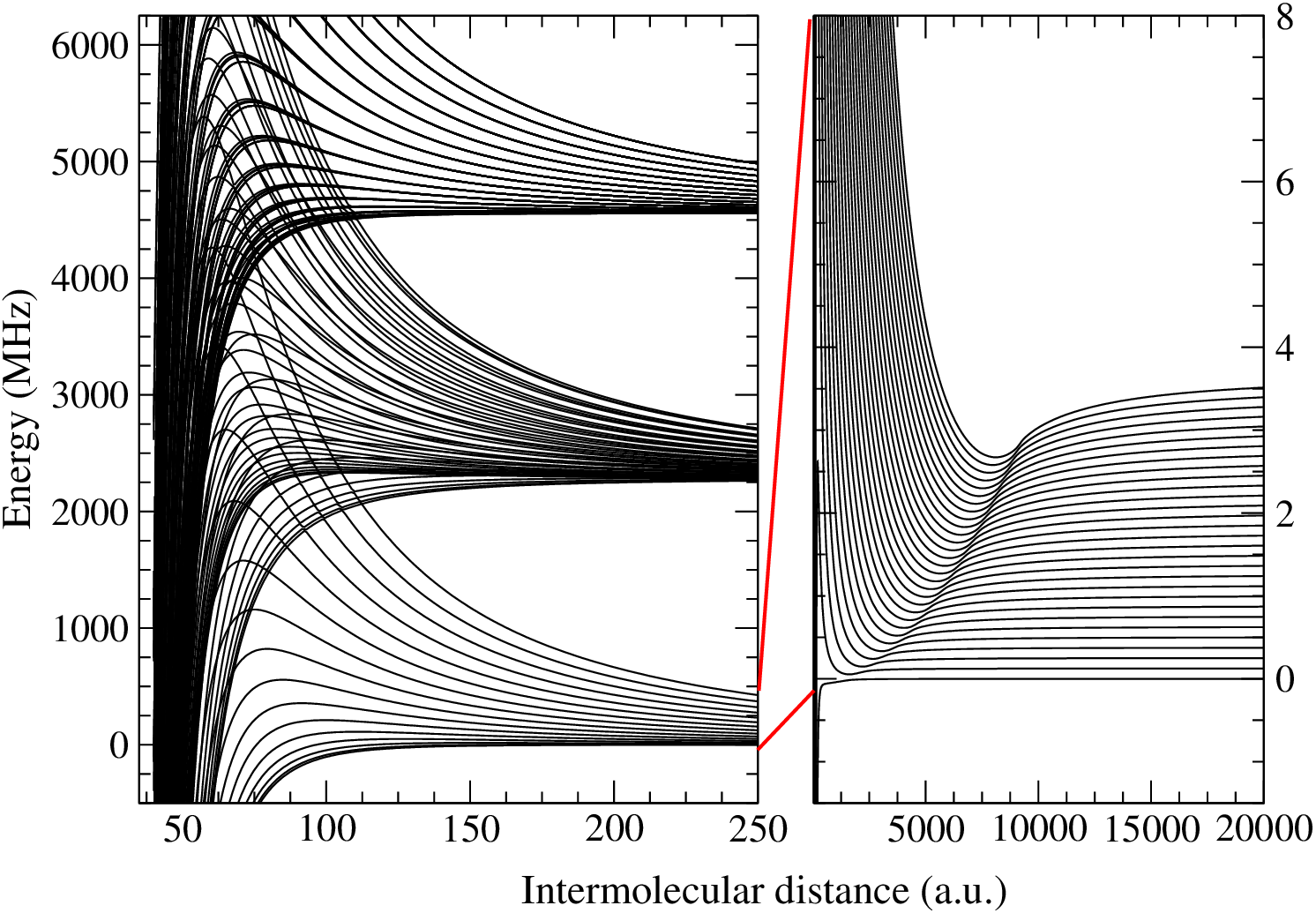}}
\caption{\label{fig:krb-adiab}  
Adiabatic potential curves for KRb + KRb system in the presence of
a static electric field of $5$~kV/cm parallel to the trap axis.
Left panel shows the short range intermolecular-distance domain,
right panel the intermediate range distance domain. Transition from a
dipolar-dominated system to a 1D confined system is indicated by the
``ridge'' visible at intermediate distances in the right panel.
 electric field of $5$~kV/cm is parallel to the trap axis.
}
\end{center}
\end{figure}

This intuitive picture is confirmed by the analysis of the adiabatic
potentials, obtained by diagonalizing the total interaction potential
at each value of the interparticle distance $R$. Note that at parallel
configuration the projection of the orbital angular momentum on the
trap axis is conserved and it has been fixed to $M_l=0$. One can observe in
Fig.~\ref{fig:krb-adiab} that the lowest adiabatic curve, the one that to
first approximation controls the dynamics, presents no potential barrier
preventing the molecules from reaching the short-range reactive region.
On the converse, the more excited adiabatic potentials present at short
range $\sim50a_0$ a barrier arising from the centrifugal potential.
The rotational degrees of freedom correspond to the excited thresholds
$\sim$GHz in the leftmost panel of Fig.~\ref{fig:krb-adiab}. Each
rotational manifold presents in turn a finer energy structure due to
the transverse harmonic trap levels with equal spacing $h \nu_{\perp}$;
See rightmost panel of Fig.~\ref{fig:krb-adiab}.

\begin{figure}[p]
\begin{center}
\resizebox{1.0\columnwidth}{!}{\includegraphics{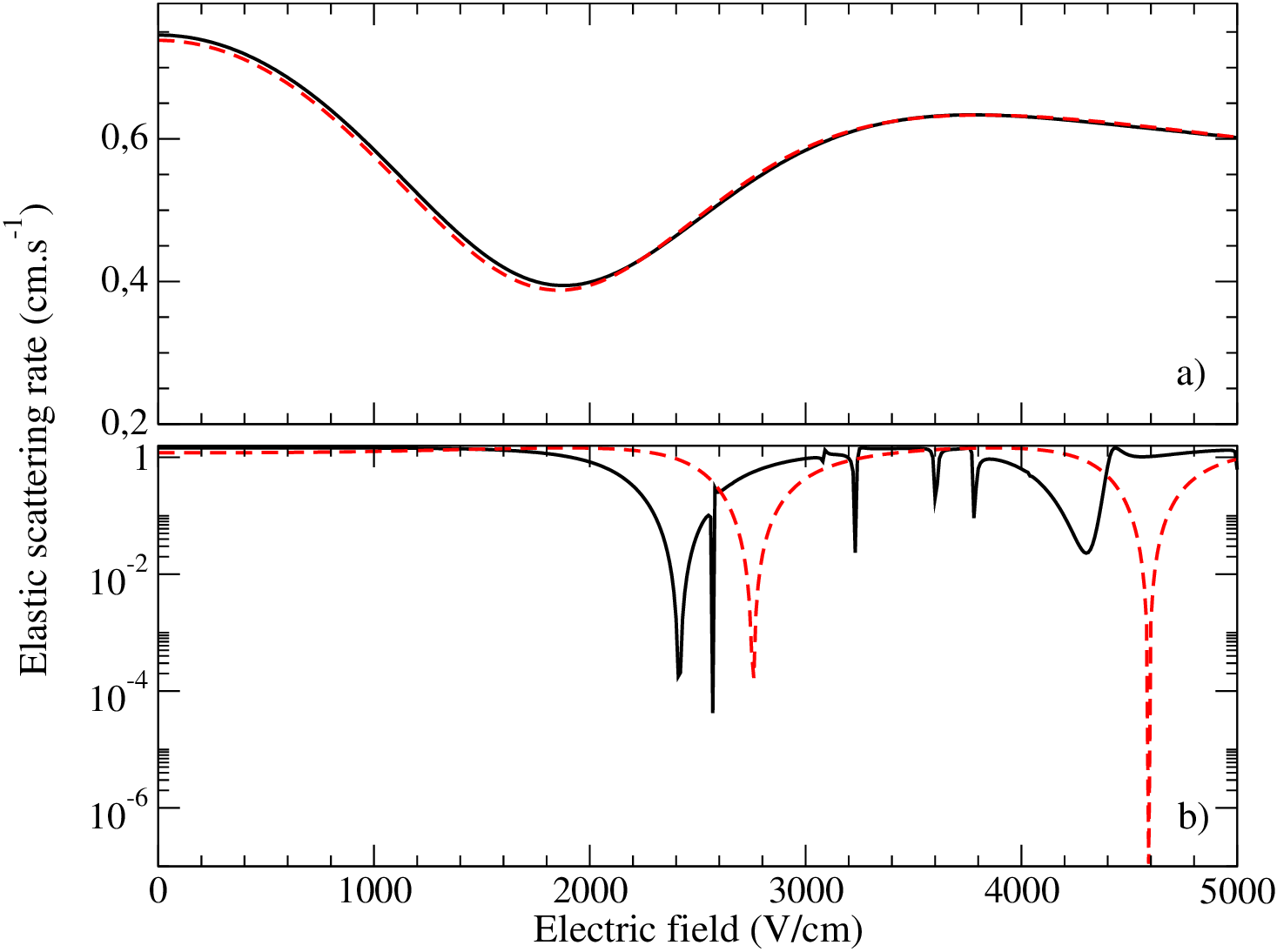}}
\caption{\label{fig:krb}  (color online) 
Elastic collisions rate between two KRb molecules, described as rigid
rotor (full line) or fixed dipole (dashed line), in a quasi-1D geometry
as a function of the amplitude of the electric field, oriented parallel
to the trap axis. Upper panel : rates computed with a unitary loss at
short-range. Bottom panel : rates computed using the Dirichlet boundary
condition. 
}
\end{center}
\end{figure}

Coming to the dynamics, we find that for the universal reactive model
the effect of the rotationally closed channel is minor. In fact, as
shown in the upper panel of Fig.~\ref{fig:krb}, the elastic collision
rates computed within the rigid rotor and the fixed dipole models are
essentially identified.
The difference is more pronounced when taking the Dirichlet boundary
condition, as illustrated in the lower panel. When the scattering phase
crosses a multiple of $\pi$, a broad zero crossing is observed in both
the rigid-rotor and the fixed-dipole calculations. This happens near
2500~V/cm in both models, with a relative shift of few hundreds~V/cm.
A second zero occurs near 5000~V/cm. By analogy with the Ramsauer-Townsend
effect, which occurs as a function of collision energy, in the following
we will briefly refer to such zeros as Ramsauer minima.

Most importantly, for rigid rotor collisions we observe resonance effects,
manifesting themselves as a series of narrow features superimposed to the
slowly varying background. As in the case of atoms, such resonances can
in principle be used to control the dimer-dimer scattering properties
through an applied field. In this work we do not attempt a precise
resonance assignment, which would require for instance quasi-bound-state
or reduced adiabatic calculations. However, the fact that such resonances
are absent in the fixed-dipole calculation strongly suggests the ones in
the figure are of rotational origin, {\it i.e.} they can be assigned to
some potential curve correlating with rotationally excited molecules. A
related study of true (as opposed to quasi-) bound states of molecules
in a quasi-1D geometry can be found in~\cite{2018-AD-PRA-063618}.

\begin{figure}[p]
\begin{center}
\resizebox{1.0\columnwidth}{!}{\includegraphics[width=\columnwidth]{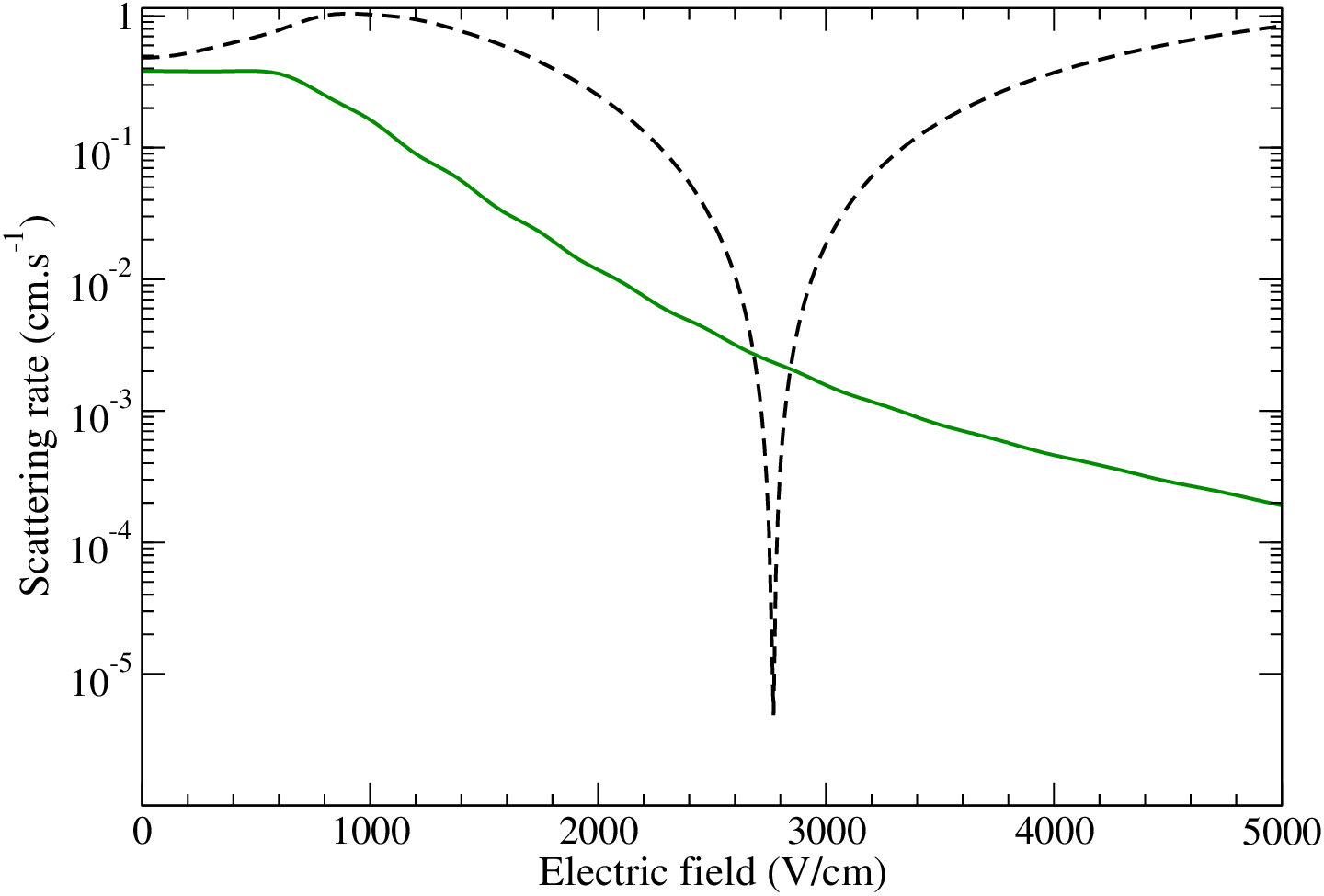}}
\caption{\label{fig:lirb-perp} (color online) Reactive (full line) and elastic (dashed line) collisions rate between two
LiRb molecules in a quasi-1D geometry as a function of the amplitude of
the electric field, oriented perpendicular to the trap axis.
Results are for a rigid rotor model, indistinguishable on the figure scale
from the ones for fixed dipoles (not shown).
}
\end{center}
\end{figure}

We now focus on a system where the electric field is perpendicular to the
trap axis. It can be expected that in this configuration and with a strong
enough induced permanent dipole moment the diatomic molecules will repel
each other at long range~\cite{2011-AS-JPB-235201} and can be protected
against short-range reactive collisions. Figure~\ref{fig:lirb-perp}
shows the elastic and reactive collision rates for LiK
molecules calculated with the fully absorbing boundary condition. Results obtained
in the fixed-dipole approximation are virtually indistinguishable from
the rigid-rotor model and are not shown. In particular, the reactive
rate is suppressed by about three orders of magnitude in the considered
range of $\cal E$, confirming that the shielding phenomenon is robust versus rotation.
Also note from the figure that the elastic rate presents a
much more pronounced minimum in the elastic cross section as compared to
KRb. This happens since, due to the small magnitude of the reactive rate for large
electric field, LiK collisions are essentially elastic and the minimum
is not quenched by inelastic processes as it happens for KRb.

Not all molecules have indeed sufficiently strong dipole moment and thus
dipole-dipole repulsion to suppress reactive processes. In general,
molecules with a large intrinsic dipole moment and a small rotational
constant are more polarizable, the induced electric dipole in the
laboratory frame is larger and thus the shielding is more effective
in these systems for a given electric field.  To confirm this trend,
we have carried out sample calculations with the trapping parameters
from Tab.~\ref{tab:data} corresponding to intermediate confinement,
and found that the only molecules not having a strong enough permanent
dipole moment to obtain significant shielding effects are LiNa and KRb.

\begin{figure}[p]
\begin{center}
\resizebox{1.0\columnwidth}{!}{\includegraphics[width=\columnwidth]{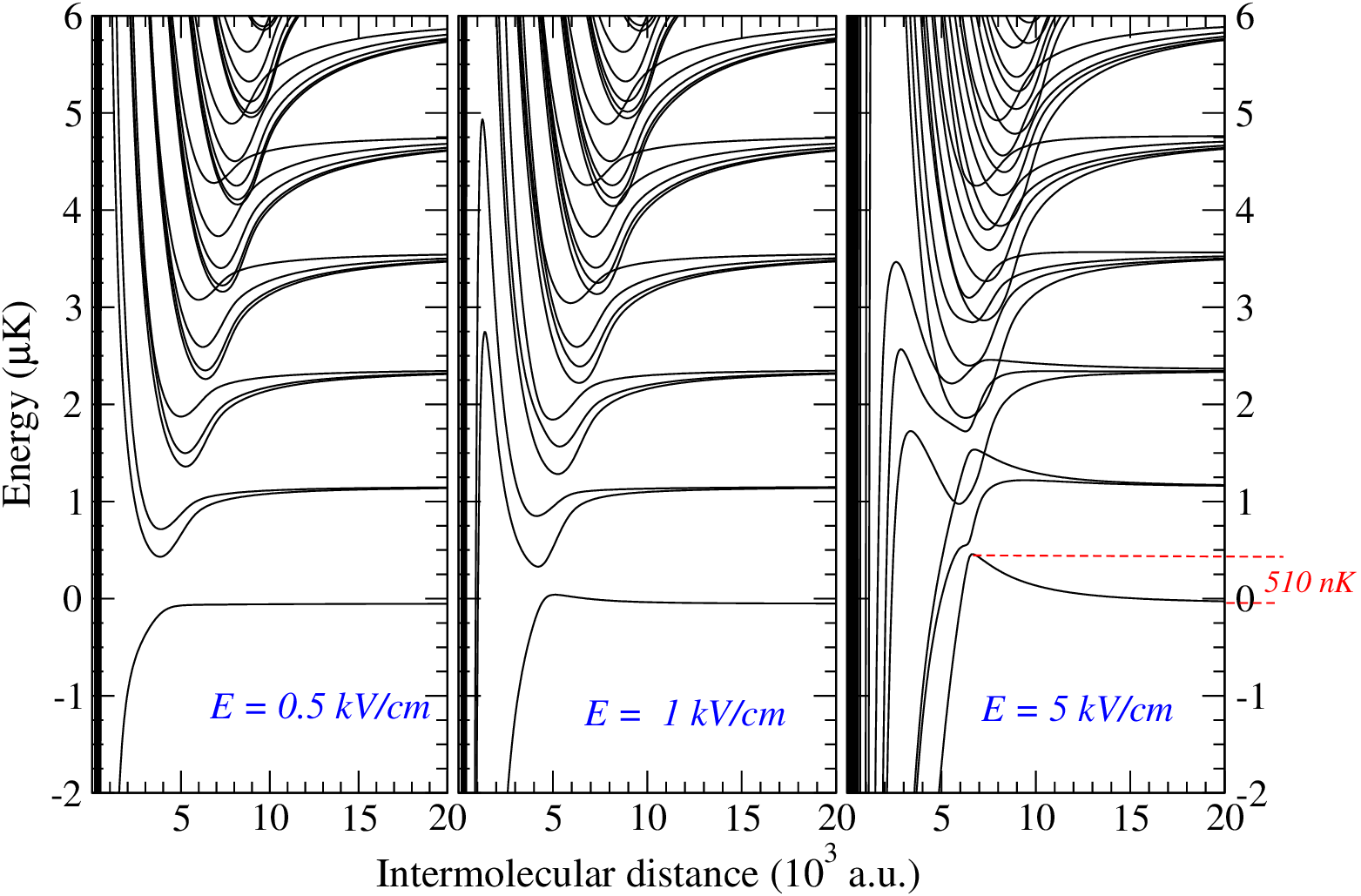}}
\caption{\label{fig:narb-adiab} Adiabatic potential curves for
NaRb + NaRb system. Left panel with a weak static electric field of
$0.5$~kV/cm, middle panel with a field of $1$~kV/cm and right panel with
a strong field of $5~$kV/cm.}
\end{center}
\end{figure}

We now consider non-reactive species, taking the NaRb dimer as an
example. The long range adiabatic curves of the NaRb-NaRb tetrameric
taken for different amplitudes of the static electric field are shown
in Fig.~\ref{fig:narb-adiab}. Each adiabatic potential correlates
asymptotically with an energy level $h \nu_{\perp} (n+1)$ of the
isotropic transverse harmonic oscillator with principal quantum number
$n$ and degeneracy $(n+1)$. At shorter distance, the dipolar interaction
becomes significant and breaks the isotropy of oscillator. The dipolar
interaction contributes for instance an energy of $d^2/R^3$ for molecules
oscillating in the plane perpendicular to the dipoles and $d^2/R^3 \left[
1-3 \cos^2(\theta) \right]$ for molecules in the plane containing the
dipoles and the trap axis, with $\theta$ the angle between $\mathbf R$
and $\mathbf d$. This anisotropy leads to the lifting of the asymptotic
degeneracy clearly visible in the three panels of the figure as the
intermolecular distance decreases.

Moreover, as expected at perpendicular configuration, a barrier to
reaction is formed in the lowest adiabatic potential as the amplitude
of the field increases. To experience a significant short-range
dynamics the molecules would need to tunnel through this barrier,
which for instance at $5$~kV/cm has height of 510~nK; See rightmost
panel of~Fig.~\ref{fig:narb-adiab}. Under this field-induced shield it
is interesting to compare the dynamics of a collision with a collision
energy well below the maximum of the barrier ($50$~nK), and slightly
over the top of the adiabatic potential barrier ($600$~nK).

\begin{figure}[p]
\begin{center}
\resizebox{1.0\columnwidth}{!}{\includegraphics[width=\columnwidth]{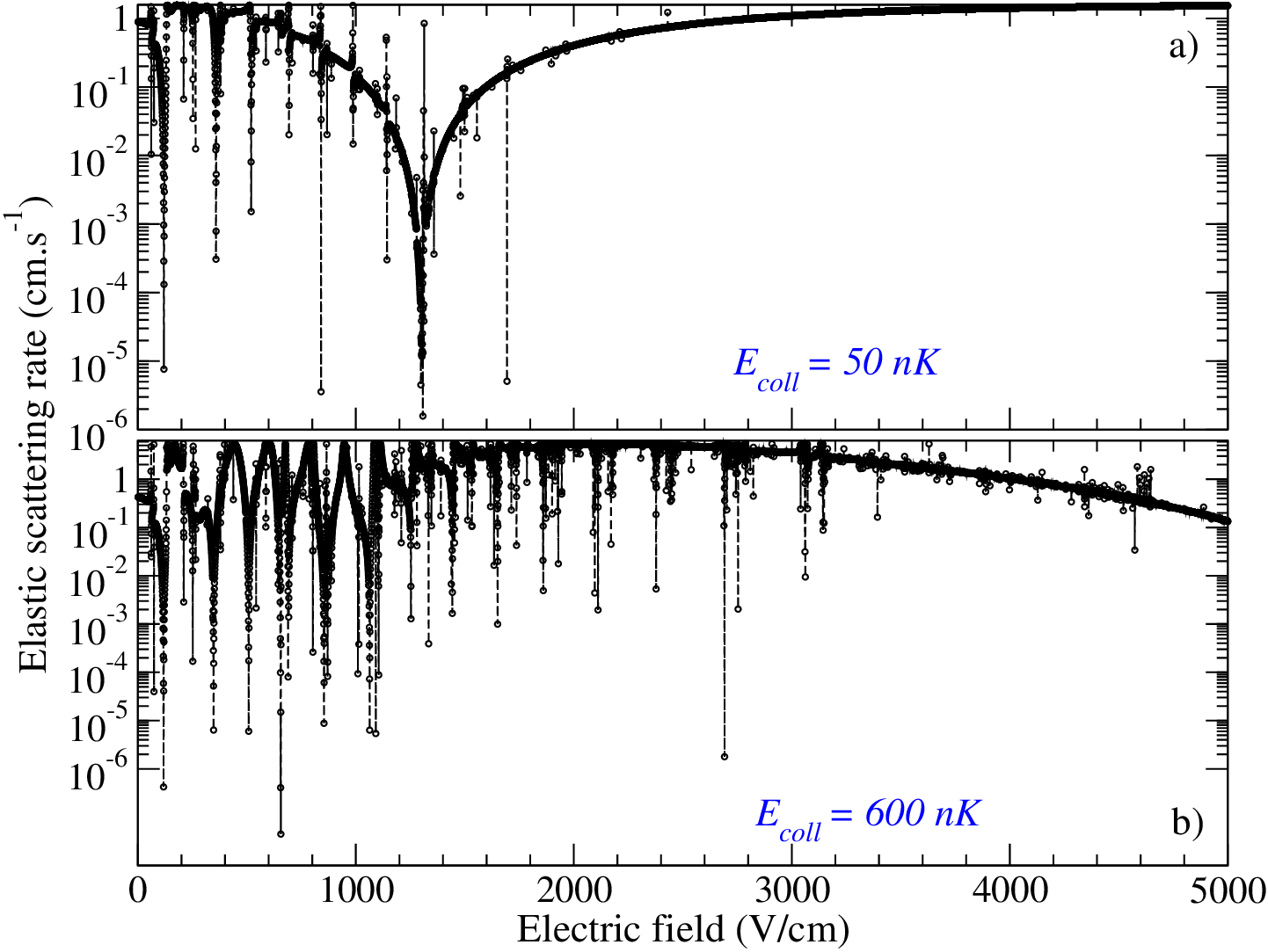}}
\caption{\label{fig:narb-temp} 
Elastic collision rate for NaRb + NaRb collisions as a function of the
amplitude of the electric field, perpendicular to the trap axis. Collision
energy is 50 and 600~nK for lower and upper panel, respectively. Dirichlet
boundary condition is imposed at short range.
}
\end{center}
\end{figure}

Figure~\ref{fig:narb-temp} shows the results of calculations performed
with the Dirichlet boundary condition. With reference to the upper
panel of the figure, one may once again remark a Ramsauer minimum near
1200~kV/cm at the lowest considered collision energy. The minimum
shifts at larger electric fields, outside the range of the bottom
panel of the figure, at larger collision energy. In fact, quite
generally, a potential has less influence on faster particles and
in order to have the accumulated phase shift go through $\pi$ larger
values of $\cal E$ are needed. The main scattering feature is the dense
spectrum of overlapping resonances observed at both considered collision
energies. Such resonances arise from the coupling between the incoming
channel and the closed channels, either correlating with the trap or
with rotationally excited level.

Resonance observed in both panels have positions essentially independent
of collision energy, as expected since a collision energy of the order of
$nK$ is essentially negligible on the scale of the Stark shift $-d \cal
E$. Note however that both the lineshape and the resonance widths vary
significantly between the two panels of Fig.~\ref{fig:narb-temp}. Most
importantly, at low collision energy (upper panel) resonance effects
tend to be washed out in particular at strong fields, say above ${\cal E}
\sim 2000$~kV/cm, since the adiabatic barrier becomes increasingly high
and broad. In other terms, the presence of the barrier tends to keep
the molecules far apart and prevents resonance effects, which are due to
a {\it short-range} coupling between the open and the closed collision
channels, from occurring. At larger collision energy barrier tunnelling
is more effective and resonances only begin to disappear near the upper
limit of the figure ${\cal E} \sim 5000$~kV/cm .

As conjectured in~\cite{2012-MM-PRA-062712,2013-MM-PRA-012709}
resonant population of dense quasi-bound states increases
the collision lifetime along with the probability of loss
of the untrapped complex and of inelastic recombination via
collision with a third body. These phenomena are suspected to be a
significant limiting mechanism to the lifetime of ultracold quantum
gases~\cite{2014-PKM-PRL-255301,2015-JWP-PRL-205302,2016-MG-PRL-205303}.
According to the present results, resonances can be suppressed in 1D
geometries. Confinement combined with a strong static electric field and
low temperatures should thus allow one to shield non-reactive molecules
from complex-mediated collisions in addition of shielding reactive molecules.

\begin{figure}[p]
\begin{center}
\resizebox{1.0\columnwidth}{!}{\includegraphics[width=\columnwidth]{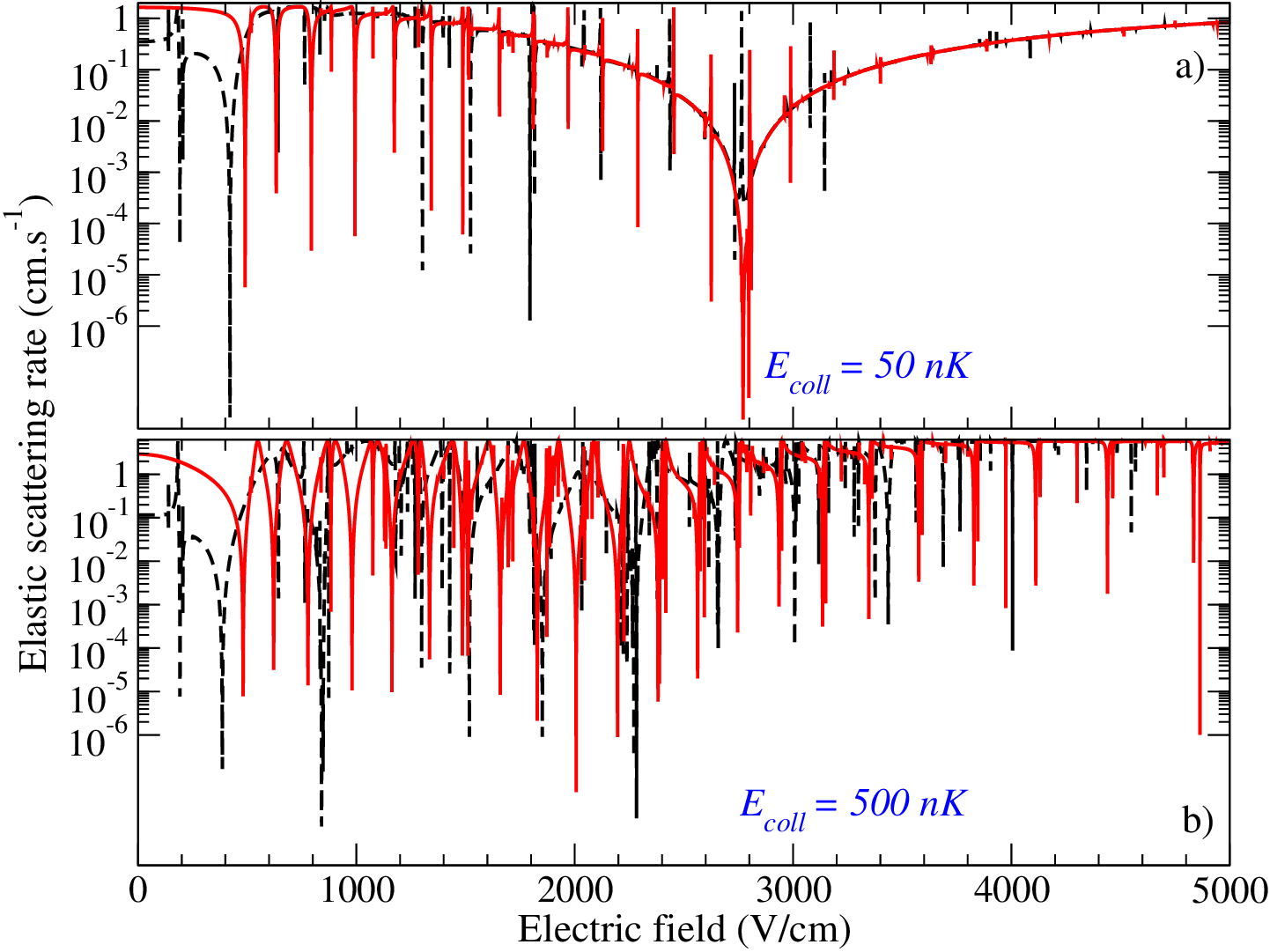}}
\caption{\label{fig:LiRb-temperature} (color online) Same as
Fig.~\ref{fig:narb-temp} but for LiRb + LiRb collisions. Solid red
lines correspond to the rigid rotor model, black dashed lines to the
fixed dipole one. Plot is for 
transverse confinement $\nu_{\perp}=2$~kHz and Dirichlet
boundary condition.}
\end{center}
\end{figure}

In order to confirm such conclusion on a different molecular species, we
consider a non-reactive LiRb model; see Fig.~\ref{fig:LiRb-temperature}.
As for NaRb, at the smaller collision energy of the upper panel,
the resonance width strongly decreases with $\cal E$, resonances first
tend to become non-overlapping and their influence on the background
cross section finally vanishes. The presence of a Ramsauer minimum,
its shift, and the behavior of the resonance width with increasing
collision energy (lower panel) follow the same trend observed in NaRb.

Figure~\ref{fig:LiRb-temperature} also shows an interesting comparison
with the fixed-dipole approximation. Note that as expected the density
of resonant features is larger for the rigid-rotor model due to the
additional rotational degrees of freedom of the dimers. However, the
density of resonances observed in the fixed-dipole model, thus purely due
to trap confinement and tuned via the Stark energy shift, is significant
in this system. Shielding becomes otherwise effective at comparable
values of the electric field at each considered collision energy.  It is
worthwhile to compare the upper panel of Fig.~\ref{fig:LiRb-temperature}
with Fig.~\ref{fig:lirb-perp} both referring to the same molecular system
and physical parameters, but with different (non-reactive {\it
vs} reactive) boundary conditions. Such comparison makes clear the
fact that the scattering rate for the {\it reactive} LiRb model in
Fig.~\ref{fig:lirb-perp} presents virtually no structure due to
the absence of electrically tuned quasi-bound states 
but rather because resonance effects are strongly quenched by
fast reactive decay.

Let us finally consider the effect of confinement. On physical grounds,
for dipoles perpendicular to the axis, stronger transverse confinement
should further help preventing particles from approaching. In fact, 
dipoles in side-by-side configuration repel, and can only get close by
moving away from the trap axis towards an attractive head-to-tail configuration.
In terms of adiabatic potentials, this means that the barrier
in the lowest adiabatic curve will be stronger at a given electric field
for a more confining than for a looser harmonic trap.

\begin{figure}[p]
\begin{center}
\resizebox{1.0\columnwidth}{!}{\includegraphics[width=\columnwidth]{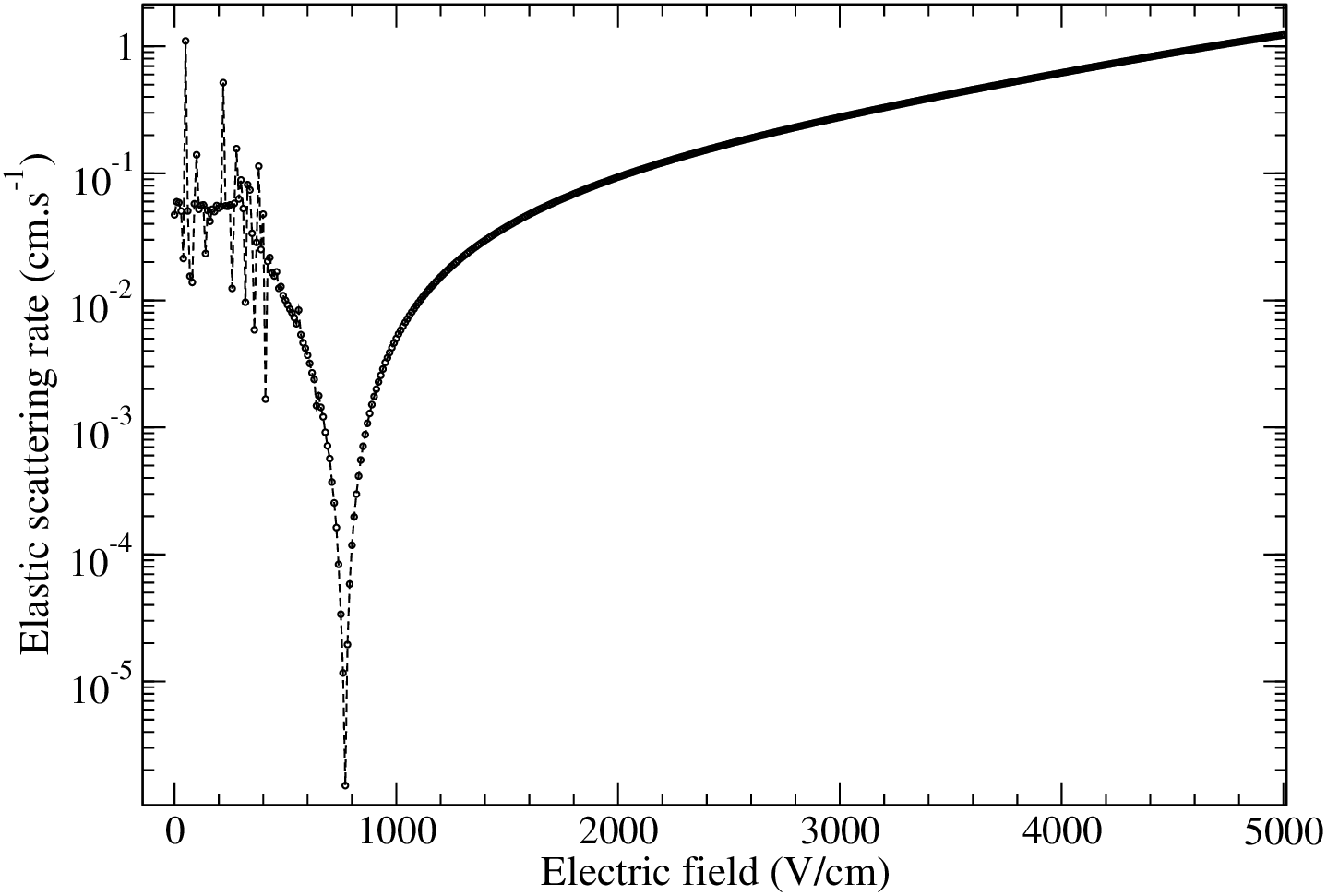}}
\caption{\label{fig:LiRb-strong} 
Elastic collision rate for LiRb + LiRb collisions as a function of
the electric field amplitude for the rigid rotor model. Plot is for tight
transverse confinement with $\nu_{\perp}=100$~kHz and Dirichlet
boundary condition. 
}
\end{center}
\end{figure}

This picture is confirmed by Fig.~\ref{fig:LiRb-strong}, that shows
the elastic collision rate for a non-reactive LiRb + LiRb collision
model in a strong confinement regime. The resonance spectrum appears
relatively sparse even at relatively low electric fields, meaning again
that coupling between the open channel and quasi-bound states is weak
since particles are prevented from reaching the short-range coupling
region.  This result extends to non-reactive systems the conclusion of
Ref.~\cite{2011-AS-JPB-235201}, that demonstrated within the fixed-dipole
approximation a more marked suppression of reactive processes occurring
in tight traps. As expected, inclusion of rotation does not seem to
change this general conclusion, that basically depends on the presence
of a potential barrier formed at long-range, in a region where the
fixed-dipole approximation is an excellent one.

Overall, it seems possible to find conditions for stabilizing the
gas against inelastic processes of both reactive or complex-mediated
nature to the extent that a suitable balance between low temperatures,
confinement and electric field intensity is found.

\section{Conclusions}

We have presented a rigid-rotor model to study identical polar molecule
collisions in a quasi-1D optical trap. Collisions of reactive molecules
in the rotational ground state are well described in the fixed-dipole
approximation. The present calculation confirms that for sufficiently
strong induced dipole moments electrostatic repulsion between dipoles
perpendicular to the trap axis leads to suppression of the reactive rates.
For non-reactive molecules, the rotational degrees of freedom result
in an increased density of Fano-Feshbach resonances. We demonstrate
that the resonance widths decrease and resonance spectra dramatically
decongest for increasing induced electric dipole moment of the molecules.
This effect could be exploited to decrease the collision lifetime 
and possibly to suppress harmful processes due to the formation
of an intermediate long-lived complex.

In perspective, it can also be interesting to model experiments where collision
dynamics has been studied in the presence of an additional optical lattice
along the axis of the tube~\cite{2012-AC-PRL-080405}. This computational
task could be accomplished for instance by combining the present 3D
solution strategy and the asymptotic reference Bloch functions constructed
in~\cite{2016-HT-PRA-032703}. The effect of hyperfine interactions could
also be included in the model to various levels of approximation.

\begin{acknowledgments}
This work is supported by the Agence Nationale de la Recherche (Contract
COLORI No. ANR-12-BS04-0020-01).

\end{acknowledgments}

\section*{References}

\end{document}